\documentstyle[epsf,epsfig,preprint,aps]{revtex}
\draft \preprint{SNUTP 01-023}
\begin{document}
\title{\Large\bf Late decaying axino as CDM and its lifetime bound}
\author{Hang Bae Kim$^{\it a}$\footnote{h.kim@lancaster.ac.uk} and
Jihn E. Kim$^{\it b}$\footnote{jekim@phyp.snu.ac.kr}}
\address{$^{\it a}$Department of Physics, Lancaster University,
Lancaster LA1 4YB, UK} \address{$^{\it b}$Department of Physics
and Center for Theoretical Physics, Seoul National University,
Seoul 151-747, Korea} \maketitle
\begin{abstract}
The axino with mass in the GeV region can be cold dark matter(CDM)
in the galactic halo. However, if R-parity is broken, for example
by the bilinear terms $\mu_\alpha$, then axino($\tilde a$) can decay to
$\nu+\gamma$. In this case, the most stringent bound on the axino
lifetime comes from the diffuse photon background and we obtain
that the axino lifetime should be greater than $3.9\times
10^{24}\Omega_{\tilde a}h$ s which amounts to a very small
bilinear R-parity violation, i.e. $\mu_\alpha<1$ keV.  This
invalidates the atmospheric neutrino mass generation through
bilinear R-violating terms within the context of axino CDM.

\noindent [Key words: dark matter, axino, R-parity violation,
detection of CDM]
\end{abstract}


\newpage

\def\smn{$\sigma_{\mu\nu}$}
\def\p{\partial}

The observed rotation curve of the halo stars~\cite{halo} requires 
to fill the galactic halo with cold dark matter(CDM) 
such as axion~\cite{ksvz,dfsz,dmaxion}, the lightest supersymmetric
particle(LSP)~\cite{dmlsp}, the axino LSP~\cite{ckr,ckkr}, and 
wimplzilla~\cite{zilla}. The axion
is motivated from the solution of the strong CP problem {\it a la}
Peccei and Quinn(PQ) \cite{pq}, which is required to be very
light~\cite{axionrv}. The
LSP is motivated from {\it R-parity conservation} in the
supersymmetric solution of the gauge hierarchy problem, where
R-parity is defined as $(-1)^{3B+L-2S}$. For the LSP
to be CDM, its mass is around 100 GeV \cite{lspbound}. The
wimpzilla is $10^{12-13}$ GeV stable particle.

The possibility of axino dark matter, which is our interest in
this paper, has been suggested from time to time as the hot DM,
the warm DM and the cold DM possibilities \cite{rtw}.
Theoretically, it arises in SUSY theories with a spontaneously
broken PQ symmetry. In supergravity it is
necessary to have a period of inflation to dilute the very weakly
interacting gravitinos. But it is thermally produced in
significant numbers even after the inflation, which requires a low
reheating temperature, $\leq 10^9$ GeV \cite{reheat}. A similar
study for axino requires a much lower reheating temperature of
order TeV in which case O(GeV) axino mass 
is allowed. Then axino can be a DM candidate \cite{ckr,ckkr}.
Here, we focus our attention on this CDM axino. Since 
there is no reliable constraint on the axino mass~\cite{ckn,ckkr},
the axino is assumed to be the LSP. Then, the most important question
is whether the LSP(axino) is absolutely stable (due to R-parity
conservation) or unstable. Since the stable axino 
case has been extensively studied \cite{ckkr}, we
restrict our attention on the R violating case.

If a CDM candidate is proposed, it is of utmost importance to
devise a scheme to prove its existence experimentally as in the
cases of the very light axion \cite{sikivie,hong} and the LSP
\cite{dmlsp,lspbound}. The axino CDM lacks this kind of possible
detection mechanism due to its extremely weak interaction strength
if R-parity is conserved. However, if R-parity is broken, it may
be possible to detect its decay products. Therefore, it is
worthwhile to study possible decay mechanisms of CDM axino. 
The detection possibility by the axino decay relies on the axino 
lifetime around $> 10^{13}$ s after the galaxy formation era.

The R-parity conservation seems to be an attractive proposal for
proton stability. However, R-parity is not dictated from any deep
theoretical principle. For example, if there
exists an $SU(3)\times SU(2)\times U(1)$ singlet superfield $N$
which is needed from the see-saw mechanism $L_\alpha NH_2$ where
$L_\alpha$ is the lepton doublet of the $\alpha$-th family and
$H_i$ is the Higgs doublet ($i=1,2$), we may write a
renormalizable superpotential, $W_N=mN^2+fN^3.$ From the coupling
to the observable sector fields for the Dirac mass coupling, $N$
is required to carry --1 unit of the R-parity quantum number. But
this R-parity is broken by the $N^3$ term. Also, $N$ obtains a
vacuum expectation value $-2m/3f$ which is too large for neutrino
phenomenology. Namely, R-parity conservation is not guaranteed
{\it a priori} at the SM level. Thus, if a singlet $N$ is
introduced, one should impose (at least an approximate) R-parity
conservation, namely we should impose $f=0$ in this example. If
R-parity is broken, it must be done so very weakly. In this
paper, for simplicity of the discussion, we restrict our attention 
on the bilinear R-violating terms,
\begin{equation}
\mu_\alpha L_\alpha H_2\label{bilinear}
\end{equation}
where $\alpha=1,2,3$. $\mu_\alpha$ is bounded by the eV order
neutrino mass. Without an explicit statement, this bound applies
to the heaviest SM neutrino, presumably the tau neutrino. With the
R-parity violation, the $\nu_\tau$ mass arises from the see-saw
type diagram with an intermediate zino line with two insertions of
the R-parity violating $\langle \tilde \nu\rangle$. Also, it can
get a contribution from the intermediate $\tilde H_2^0$ line with
two insertions of $\mu_3$. These give similar conclusions and we
discuss the $\mu_3$ case for an explicit illustration. Then,
$\mu_3$ is bounded as
\begin{equation}
|\mu_3|\le M^{1/2}_{\tilde H_2^0,TeV}\ {\rm MeV}\label{mu3}
\end{equation}
where $M_{\tilde H_2^0,TeV}$ is neutral Higgs mass in units of
TeV. In this paper, we introduce dimensionless numbers: for a
small coupling $\epsilon$, $\epsilon_{-n}$ represents it in units
of $10^{-n}$, for MeV order masses $m_{MeV}$ in units of MeV, for
GeV order masses $m_{GeV}$ in units of GeV, for large mass
$m_{[n]}$ in units of $n$ GeV, and for super large mass $F_{12}$
in units of $10^{12}$ GeV. 

The bilinear R-violating parameters
have been extensively discussed in regards to the neutrino
oscillation \cite{chunR} within the above bound (\ref{mu3}). In
this paper, we will draw a conclusion that this is not consistent
with the CDM axino, already from the observed diffuse gamma ray
background.

With the bilinear R-parity violation, we expect the following
decay modes of the axino
\begin{eqnarray}
&\tilde a\rightarrow \nu+\gamma({\rm or\ }l^+l^-),\ \ \tilde
a\rightarrow \nu + a,\ \ \tilde a\rightarrow\tau^++\pi^-,\ {\rm
etc.}
\end{eqnarray}
where $m_{\tilde a}>m_\tau+m_\pi$ is assumed.

To estimate the partial decay widths, let us assume the following
R violating axino interaction
\begin{equation}
{\cal L}_{\tilde a-{\rm decay}}= \epsilon_0\ \phi \tilde a \psi\
,\ {\rm or}\ \ i\epsilon_1\frac{\alpha_{em}}{F_a}F^{\mu\nu}\tilde
a\gamma_5[\gamma_\mu,\gamma_\nu]\psi\label{effint}
\end{equation}
where $F_a$ is the axion decay constant(including the division by
the domain wall number $N_{DW}$), $F_{\mu\nu}$ is the field
strength of a spin-1 field $A_\mu$, $\phi$ is a scalar field and
$\psi$ is a fermion field (Dirac or Majorana field). Then, the
lifetime of axino becomes
\begin{equation}
\tau_{\tilde a}= n_\psi m^{-1}_{\tilde a,GeV}\cdot
\left(1.32\epsilon_{0,-11}^{2}\cdot P_0,\ {\rm or}\ \ 2.57\times
10^{-5} \epsilon_1^{2}F_{a,12}^{-2}m^2_{\tilde a,GeV}\cdot P_1
\right)^{-1}\ [{\rm sec}] \label{life}
\end{equation}
where $n_\psi=1,2$, respectively for the Dirac and Majorana
$\psi$, we neglected $m_\phi$, and $P_{0,1}$ are phase space
factors. For a massive final fermion, $
P_{0,1}^{-1}=(1-{m_\psi^2}/{m^2_{\tilde a
}})^{-1}(1+{m_\psi}/{m_{\tilde a}})^{-2}.$ Let us proceed to
discuss several possibilities of O(GeV) axino decay.

Firstly, for the $\tilde a\rightarrow \nu+a$ decay, we note that
the axion multiplet couples to the standard model chiral fields,
below the PQ symmetry breaking scale, as $\exp({iQA/F_a})W_Q$
where $W_Q$ carries $-Q$ units of the Peccei-Quinn(PQ) charge, and
$A$ is the axion supermultiplet. The heavy quark axion models
\cite{ksvz} do not allow these tree level couplings since the SM
fields are neutral under the PQ symmetry, but the lepton-coupling
type models \cite{dfsz} can lead to this kind of couplings. Our
interest is on the superpotential for the axino-neutrino coupling,
$ m_\nu\nu\nu\exp({2iQ_\nu A/F_a})$ where $Q_\nu$ is the PQ charge
of the neutrino, and $\mu_\alpha L_\alpha H_2 e^{iQA/F_a}$.
Setting $Q_\nu=1/2, Q=1/2$, we obtain $W=m_\nu\nu\nu e^{iA/F_a}$
and $\mu_\alpha L_\alpha H_2 e^{iA/F_a}$ from which we obtain the
relevant terms for the axino decay $ -(m_\nu A^2/ 2F_a^2)\nu \nu$
and $-(\mu_\alpha A^2/2F_a^2)L_\alpha H_2$, respectively. The
R-parity violation by the vacuum expectation value of a sneutrino,
$v_{\tilde \nu}\equiv\langle\tilde \nu\rangle$, or by $\mu_\alpha$
allow the following Yukawa couplings
\begin{equation}
{\cal L}_{\tilde a\rightarrow \nu a}=\left(\frac{m_\nu
v_{\tilde\nu}}{F_a^2},\ {\rm or\ }\frac{\mu_\alpha v_2}{F_a^2}
\right)\times a \nu \tilde a\label{nudecay}
\end{equation}
from which we estimate $\epsilon_0\simeq 10^{-33} m_{\nu,
eV}v_{\tilde\nu, GeV}/F^2_{a,12}$ and $10^{-25} \mu_{\alpha, MeV
}v_{2,[100]}/F^2_{a,12}$, respectively, and $v_2=\langle
H_2^0\rangle$. Thus, in view of Eq. (\ref{life}) $\epsilon_0$ is
too small (i.e. $\tau_{\tilde a}\sim 10^{28}$ s) and the decay
mode $\tilde a\rightarrow\nu+a$ is not important cosmologically.

Second, the decay $\tilde a\rightarrow \tau+\pi^+$ occurs through
the bilinear R-parity violating term $\mu_\alpha L_\alpha H_2$
with $\alpha=3$, given in Eq. ({\ref{bilinear}}), which allows the
mixing of $\tilde\tau$ and $H_1^-$. Then, the coupling
$(m_\tau/F_a)\tilde\tau^+\tau \tilde a$ and the Yukawa coupling
$\sim (m_d/v_1)qd^cH_1$ give an effective interaction
(\ref{effint}) with
\begin{equation}
\epsilon_0=\frac{m_\tau f_\pi
m_\pi^2}{F_av_1}\frac{1}{M^2_{\tilde\tau}}\Delta
M^2_{RPV}\frac{1}{M^2_{H_1^-}}
\end{equation}
where $\Delta M^2_{RPV}$ is the $\tilde \tau$--$H_1^-$ mixing
parameter. The effective interaction (\ref{effint}) with
$\psi=\tau$ and $\phi=\pi^+$ arises from the tree diagram with the
$\tilde\tau-H^-$ mixing insertion in the intermediate scalar
propagator with the four external fermions, $\tilde a, \tau, \bar
d,$ and $u$. In estimating $\epsilon_0$, we used the PCAC relation
in obtaining the matrix element $\langle 0|\bar d_R
u_L|\pi^+\rangle\sim f_\pi m_\pi^2/m_d$. The superpartner masses
for the gauge hierarchy solution are around 100 GeV. The bound on
the stau-charged Higgs mixing parameter $\Delta M^2_{RPV}$ is
bounded from the tau neutrino mass bound, $m_\nu<1$ eV. For the
R-parity violating bilinear coupling $\mu_3 L_3H_2$, the mixing
parameter is estimated as $\Delta M^2_{RPV}= 2\mu_3^*\mu.$ Using
Eq. (\ref{mu3}), we obtain $\Delta M^2_{RPV}<2\mu_{TeV}M_{\tilde
H_2^0, TeV}^{1/2}$ GeV$^2$. Then, we estimate
$\epsilon_0<3.71\times 10^{-25}[\cos\beta]^{-1} F^{-1}_{a,12}
\mu_{TeV}M^{-2}_{H_1^-,[100]}M^{-2}_{\tilde\tau,[100]}
M^{1/2}_{\tilde H^0_{2,TeV}}$ for the $\tau\pi$ decay mode, and
the axino lifetime must satisfy a bound
\begin{equation}
\tau_{\tilde a}\ge 0.959\times 10^{27}\ [{\rm sec}]\
\frac{M^4_{\tilde\tau,TeV}M^4_{H_1^-,TeV}}{m_{\tilde
a,GeV}\mu^2_{TeV}M_{\tilde H^0_{2,TeV}}}\cdot P_0^{-1}
\label{taulowtau}
\end{equation}
where the MSSM parameter $\tan\beta=v_2/v_1$, and we assume that
$P_0$ is nonzero, i.e. the O(GeV) axino has mass $m_{\tilde a}>
1.92$ GeV.

Third, we note that the interaction $\nu\tau\bar\tau\tilde a$,
arising from stau intermediate state and R-parity violating
insertion of neutrino--$\tilde B/\tilde W$ mixing, is not
important.

Finally, we note that the decay $\tilde a\rightarrow \nu+\gamma\
({\rm or}\ l^+l^-)$ occurs through the anomaly term \cite{ckkr},
\begin{equation}
{\cal L}_{a\gamma\nu}= i\frac{c_{a\gamma\gamma}\alpha_{em}}{16\pi
F_a}\cdot\frac{c^\prime\mu_\alpha}{\mu}\cdot\bar\nu_\alpha
\gamma_5[\gamma^\mu,\gamma^\nu]\tilde a F_{\mu\nu}\label{photon}
\end{equation}
where $c_{a\gamma\gamma}$ is the axion-photon-photon coupling
which depends on models \cite{cgg}, and $F_{\mu\nu}$ is the photon
field strength, and the photino-neutrino mixing parameter
$c^\prime\mu_\alpha/\mu$ has been introduced. Note that $c^\prime$
is O($<$1) and $\mu_\alpha$ is O($<$MeV). It turns out that this
photon mode constitutes the most important contribution in the
axino decay. The gluon anomaly term can be considered, but it is
not important since we must consider intermediate gluino and
squark lines. From the interaction Eq. (\ref{photon}), we estimate
\begin{equation}
\epsilon_1= 1.99\times
10^{-7}c_{a\gamma\gamma}c^\prime\mu_{\alpha, MeV}\mu_{[100]}^{-1},
\end{equation}
giving the axino lifetime
\begin{equation}
\tau_{\tilde a}\simeq 9.8\times 10^{17}\ [{\rm sec}]\
c_{a\gamma\gamma}^{-2}c^{\prime -2}\mu_{\alpha, MeV}^{-2}
\mu^2_{[100]}F_{a,12}^2 m^{-3}_{\tilde a, GeV}.\label{taulow}
\end{equation}
The RHS of Eq. (\ref{taulow}) can fall in the cosmologically
interesting scale for $F_a$ slightly smaller than $10^{12}$ GeV
and $m_{\tilde a}$ = O(GeV).

This leads us to the estimation of the cosmological abundance of
axino. The decoupling temperature of axino is of order the PQ
symmetry breaking scale~\cite{rtw,ckkr}. Thus, for an O(GeV)
axino, inflation must end below the PQ symmetry breaking scale so
that axinos produced at the decoupling temperature is sufficiently
diluted. However, if the reheating temperature after inflation
were high enough, a significant number of axinos would have been
reproduced thermally and can constitute cold dark matter. Here, we
are interested in this thermally produced axinos after inflation.
In this scenario, the number density depends on axino mass and
reheating temperature $T_R$. For O(GeV) axino in R-conserving
theories the reheating temperature bound is $T_R\leq 100-1000$ GeV
from the condition that the thermally produced axinos do not
exceed the critical energy density as estimated in
Ref.~\cite{ckkr}. On the other hand, if there exists R-violating
terms, then the lightest neutralino decays to $l^-l^+\nu$ within 1
s for $\mu_\alpha/\mu<10^{-6}$, which occurs through the diagram
$\chi_1^0\rightarrow W^\mp \chi^\pm$ and the mixing of $\chi^\pm$
and $l^\mp$. Namely, a neutralino predominantly decays to SM
particles, which is harmless at a later epoch. Thus, to produce
O(GeV) axino copiously in R-violating theories, reheating
temperature must be raised.

In the present estimation of reheating temperature $T_R$, we
ignore the gluino decay to axino since R-parity is broken. Taking
into account the photon thermalization after the decay of MSSM
SUSY particles and $e^+e^-$ annihilation, the axino energy density
at present is
$\rho_{\tilde a}(T_\gamma)=m_{\tilde
a}(2\pi^2/45)\cdot(43/11)\cdot
T_\gamma^3Y^{TP}_{\tilde a}(T_R)
$
where the thermal production of axino $Y^{TP}_{\tilde a}$ is
calculated just from the scattering processes. Thus, representing
$\rho_{\tilde a}=\Omega_{\tilde a}\times($the critical energy
density), we have $m_{\tilde a}Y^{TP}_{\tilde a}\simeq 0.72\ {\rm
eV}(\Omega_{\tilde a}h^2/0.2)$. For $F_a\sim 10^{11}$ GeV,
$\Omega_{\tilde a}\sim 0.3$ and O(GeV) axino, $Y^{TP}_{\tilde
a}\sim 0.5\times 10^{-9}$ and we can read $T_R$ at around 200 GeV
from Fig. 1 of Ref. \cite{ckkr}. In this region, if one considers
the gluino decay in R-conserving theories, $T_R$ should be a
factor $\sim 2$ smaller. The neutralino decay in R-conserving case
is not important in this region, but can be very important for
$T_R<100$ GeV \cite{ckkr}.

The present dark matter density in the galactic halo requires a
significant amount of dark matter. In the present case, there is
the other candidate for dark matter, the axion. For the axion CDM,
we are restricted to $F_a\sim 10^{12}$ GeV. But with the CDM
axino, $F_a$ can be lower as far as $F_a>10^9$ GeV. If the axino
lifetime is of order the age of the universe, then there remains a
significant number of axinos which can be detected in experiments
in search of proton decay. The interesting decay mode for axino
detection is the $\nu\gamma$ mode.

For $10^3\ {\rm sec} <\tau_{\tilde a}<t_{\rm rec}$ where $t_{\rm
rec}$ is the time at the recombination, there can be an allowed
region $\tau_{\tilde a}>t_{\rm min}$ so that the decay products
are not copious enough to dissociate the light nuclei.
After the time of recombination, photons are not
effective to scatter off the neutral particles, and hence
$t_{min}\le t_{\rm rec}$. Since we are considering the axino
lifetime $>10^{13}$ s, the late decaying axino is safe from
destroying light nuclei.

If axino decays to $\nu+\gamma$, the underground neutrino
detectors can detect the photon. The photon energy($\equiv
E_\gamma$) of 10 GeV, i.e. axino mass of 20 GeV, is the boundary
for using different search types for the Cherenkov rings. If
$E_\gamma<10$ GeV, the Compton scattering on an atomic electron
kicks out a high energy electron whose Cherenkov radiation can be
detected. If $E_\gamma>10$ GeV, then $e^+e^-$ pair production off
nucleus dominates and the Cherenkov rings from these pair can be
detected.

From the super-K detector, one establishes the proton lifetime
bound of $10^{33}$ seconds \cite{pdecay}. These detectors use
baryons in water, $n_B=N_A/$cm$^3$ where $N_A=6.023\times
10^{23}$. On the other hand axino as CDM {\it now} has the local number
density of order $n_{\tilde a}=0.3 m_{\tilde a,GeV}^{-1}$/cm$^3$,
giving the ratio $n_{\tilde a}/n_B\sim 5\times
10^{-25}m^{-1}_{\tilde a,GeV}$. If we require the detection rate
of axino decay the same as that of proton decay, we obtain
$ (\tau_p n_{\tilde a}/\tau_{\tilde a} n_p)\simeq 1.  $
Thus, axinos are detectable at the rate of counting proton decay
debris with proton lifetime of $10^{33}$ seconds \cite{pdecay} if
\begin{equation}
\tau_{\tilde a}\simeq \frac{1.6\times 10^{16}}{m_{\tilde a,GeV}}\
[{\rm sec}].\label{tauhigh}
\end{equation}
Since we have not observed this kind of events, we obtain
$\tau_{\tilde a}>1.6\times 10^{16}m_{\tilde a,GeV}^{-1}$ s.

However, the most stringent bound comes from the diffuse gamma ray
background. The classical study on this effect has been published
more than 20 years ago \cite{dicus}.
The observed flux ${\cal F}_\gamma$ is bounded by \cite{ressell}
\begin{equation}
\frac{d{\cal F}_\gamma}{d\Omega}\le (10^{-3}\sim10^{-5})\ 
E^{-1}_{GeV}\ {\rm cm}^{-2}{\rm sr}^{-1}{\rm sec}^{-1}\label{obs}
\end{equation}
where $E$ is the decay photon energy at present.
The figure $10^{-3}$ is for the conservative bound applicable
to the whole observed range of $E$.
For $E=1{\rm MeV}\sim10{\rm GeV}$
where we are interested, $10^{-5}$ gives a good fit to the data.

On the other hand, the decay of axinos produces the diffuse photon
flux at present, for $0\le E\le m_{\tilde a}/2$
\begin{equation}
\frac{d{\cal F}_\gamma}{d\Omega}=\frac{3n_{\tilde a}}{8\pi}
\left(\frac{E}{E_0}\right)^{3/2} e^{-(E/E_0)^{3/2}}
\end{equation}
where $E_0=(m_{\tilde a}/2)(\tau_{\tilde a}/t_0)^{2/3}$ and $t_0$
is the age of the universe ($\sim 4\times 10^{17}$ s).

For $t_{\rm rec}<\tau_{\tilde a}<t_0$, most axinos have decayed
and the flux has a peak at $E=E_0$ with a value $n_{\tilde a}/8\pi e$.
Assuming the critical axino density,
we have a condition that the photons from axinos decay do not exceed
the observed flux,
\begin{equation}
(\tau_{\tilde a}/t_0)^{2/3}< 1.4\times 10^{-7}\Omega_{\tilde a}h^2
\end{equation}
which is inconsistent with the condition
$\tau_{\tilde a}>t_{\rm rec}\ (\sim10^{13}$ s).

For $\tau_{\tilde a}>t_0$, axino decays are increasing at present
and the maximum flux is
$(3n_{\tilde a}/8\pi)(t_0/\tau_{\tilde a})e^{-t_0/\tau_{\tilde a}}$ at
$E=m_{\tilde a}/2$. Comparing this with Eq. (\ref{obs}), we obtain
\begin{equation}
\label{tbound}
\tau_{\tilde a, sec}>3.9\times 10^{24-26}\Omega_{\tilde a}h.
\end{equation}
For the region satisfied by Eq. (\ref{tbound}), the bound on
$\mu_\alpha$ is very stringent, i.e. less than O($10^{2-3}$ eV),
hence the idea for neutrino mass generation via bilinear R-parity
violation is not consistent with CDM axino. A very conservative
upper bound on $\mu_\alpha$ is 1 keV.

In conclusion, we searched for the detection possibility of axinos
as CDM with $T_R\sim 200$ GeV. The diffuse gamma ray background
gives a very strong bound on bilinear R-parity violating parameter
$\mu_\alpha$. Even if $\mu_{\alpha}$ is of order keV, it can be
detected by diffuse gamma ray background observation. On the other
hand, with O(keV) $\mu_\alpha$ SUSY generation of neutrino
oscillation parameters through bilinear R-parity violation is not
achievable with CDM axino.

\acknowledgments One of us(J.E.K.) has benefitted from very useful
comments by E. J. Chun and S. B. Kim. This work is supported in
part by the BK21 program of Ministry of Education, and by the
Center for High Energy Physics, Kyungpook National University.

\end{document}